Microdosimetry of a clinical carbon-ion pencil beam at MedAustron - Part 1: experimental characterization

Cynthia Meouchi [1], Sandra Barna [2], Anatoly Rosenfeld [3], Linh T. Tran [3], Hugo Palmans [4,5], Giulio Magrin [4]

[1] Radiation Physics, Technische Universität Wien, 1040, Vienna, Austria

[2] Medizinische Universität Wien, Universitätsklinik für Radioonkologie, Vienna, Austria

[3] Center for Medical Radiation Physics, University of Wollongong, Wollongong, Australia

[4] MedAustron Ion Therapy Center, Marie Curie-St. 5, 2700 Wiener Neustadt, Austria

[5] National Physical Laboratory, Hampton Road, TW11 0LW, Teddington, UK

Corresponding author: Cynthia Meouchi; email: cynthia.meouchi@hotmail.com

## Abstract

Objective: This paper characterizes the microdosimetric spectra of a single-energy carbon-ion pencil beam with a nominal energy of 284.7 MeV/u at MedAustron using a miniature solid-state silicon microdosimeter to estimate the impact of the lateral distribution of the different fragments on the microdosimetric spectra.

Approach: The microdosimeter was fixed at one depth and then laterally moved away from the central beam axis in steps of approximately 2 mm. The measurements were taken in both horizontal and vertical direction due to the fact that the pencil beams are not radially symmetric. These measurements were performed in a water phantom at different depths. In a position on the distal dose fall-off beyond the Bragg peak, the frequency-mean and the dose-mean lineal energies, $\bar{y}_F$ and $\bar{y}_D$, were derived using either the entire range of y-values, or a sub-range of y values, presumingly corresponding mainly to contributions from primary particles.

Main results: The measured microdosimetric spectra do not exhibit a significant change up to 4 mm away from the beam central axis. For lateral positions more than 4 mm away from the central axis, the relative contribution of the lower lineal-energy part of the spectrum increases with lateral distance due to the increased partial dose from secondary fragments. The average values $\bar{y}_F$ and $\bar{y}_D$ are almost constant for each partial contribution. However, when all particles are considered together, the average value of $\bar{y}_F$ and $\bar{y}_D$ varies with distance from the axis due to the changing dose fractions of these two components varying by 30 % and 10 % respectively up to the most off-axis vertical position.

Characteristic features in the microdosimetric spectra providing strong indications of the presence of helium and boron fragments have been observed downstream of the distal part of the Bragg peak.

Significance: We were able to investigate the radiation quality as function of off-axis position. These measurements emphasize variation of the radiation quality within the beam and this has implications in terms of relative biological effectiveness.

## 1. Introduction

The use of carbon-ion beams in cancer therapy has increased lately due to its favorable depth-dose distribution with a distinct peak (the Bragg peak) and a sharp distal edge. Furthermore, carbon ions induce a higher biological response in the Bragg peak which enhances its advantage as compared to protons. The variation of this biological response needs to be modelled in dose-response calculations in clinical treatment planning calculations. Given the link between the biological response and the radiation quality, which has recently received considerable interest[1], the most efficient way of modelling this is by using the 3D dependence of the radiation quality parameters in the pencil carbon ion beam. This defines the need for characterizing this radiation quality dependence in detail.

Carbon ions exhibit less lateral beam spread than protons, but, on the other hand, undergo nuclear fragmentation reactions along their penetration path. Hydrogen, helium, lithium, beryllium, and boron fragments are produced with different ranges, populating the radiation field in a heterogeneous way. Consequently, fragmentation affects the radiation quality of the field with depth and laterally. Different groups studied the secondary beam fragments production by $^{12}$C ions [2–5] in thick water or PMMA targets using the time of flight (TOF) technique to measure the energy of the fragment particles produced at different depths. Microdosimetric evaluation of secondary beam fragments was also performed [6,7] using a tissue equivalent proportional counter (TEPC) and the TOF method to separate the spectra of each component of the charged-particle species and experimentally using monolithic ΔE-E telescope followed by comparison with Monte Carlo simulations [8]. In-phantom microdosimetric characterization of a carbon ion beam was performed using a TEPC by Gerlach *et al.*[9] and Monte Carlo simulations by Galer *et al.*[10]. Both studies also investigated the biological effect via a biological response function. In order to characterize the complex radiation field, Martino *et al.*[11] performed out of field lateral measurements at different depths using the TEPC in $^{12}$C and $^{7}$Li pencil beams showing that the radiation quality, expressed as the dose-mean lineal energy $\bar{y}_D$, steeply decreases as a function of the distance from the beam central axis to a constant value of 1-2 keV μm$^{-1}$ above 2 cm. Those studies were done using a large TEPC moved transversally to the beam central axis in steps of a few centimeters.

In this paper, we characterize the radiation quality changes by means of the microdosimetric spectra along and laterally to the beam central axis for a 284.7 MeV/u carbon ion pencil beam using a miniature silicon microdosimeter at different lateral positions with high spatial resolution. One non-trivial requirement we address is the use of detectors, setups and methodologies that are feasible for frequent use in clinical beams. This is the first time the MedAustron carbon ion pencil beams have been characterized in this way. Compared to previously published studies with a similar aim[11], we use a much smaller size microdosimeter and a more detailed characterization in steps of a few millimeters.

## 2. Materials and methods

The measurements were carried out in the MedAustron research irradiation room using a pencil beam of 284.7 MeV/u carbon ions delivered by repeating the same single spot several times to achieve a high integrated count and with the lowest achievable flux. The pencil beams in the research room have exactly the same characteristics and specifications as the clinical pencil beams in the treatment rooms. The FWHM of the pencil beam profile at the phantom surface is 6.6 mm. The data were taken during three different measurement sessions.

The so-called silicon on insulator (SOI) 3D mushroom microdosimeter developed at the Center for Medical Radiation Physics (CMRP) of the University of Wollongong, Australia has been employed for the measurements and is further referred to in the paper as the silicon microdosimeter. The silicon microdosimeter has an array of 400 cylindrical sensitive volumes (SVs) with pitch of 50 µm, each having a diameter of 18 µm and a thickness of 10 µm fabricated on high resistivity p-type silicon on insulator active layer. The total active area of the silicon microdosimeter is about 0.1 mm$^2$. Additional details can be found in references [12,13].

A schematic illustration of the set-up is shown in **Figure 1** for one specific depth, where measurements at six lateral horizontal positions along the x-axis, and at ten vertical positions along the y-axis were carried out. The silicon microdosimeter was mounted along the beam's central axis (z) at four depths in the water phantom (Water Phantom 41023 for Horizontal Beams, PTW Freiburg, Germany). The detector was placed at the desired depth, and the pencil beam was moved laterally along x- or y- axis in steps of 2 mm.

The apparatus was positioned in the water phantom using the indication of the lasers and the fiducial references on the detector holder. The uncertainty of alignement between the microdosimeter and the beam axis is estimated to be 0.6 millimeters in both directions. This accounts for the combined effect of the setup of the phantom, the slightly loose fixing of the detector to the holder, and the misalignement between the laser and the beam axis position. The total uncertainty was previously estimated in the MedAustron beam line for a different apparatus at analogous conditons[14]. The microdosimetric spectra were collected in those lateral positions at four depths along the beam central axis (see **Figure 2**). A is the position in the plateau, B is at the Bragg peak, C is at the depth $R_{55}$ corresponding with the 55% percentage depth dose value in the distal part of the Bragg peak and D is at the depth $R_{08}$ corresponding with the 8% percentage depth dose value on the distal part of the Bragg peak.

The pulse-height spectra were converted to spectra of lineal energy based on the two-step method described in Meouchi et al.[15]. The first step of this method consisted of correcting deviations from linearity of the readout electronics (pre-amplifier, shaping amplifier, and multi-channel analyzer (MCA)). The so-called double linearization was used which involves fitting two parts of the data with separate linear curves that are joint at a transition channel value (hinge point). The second step of the method concerns the calibration of lineal energy values of the spectrum using the edge technique. This refers to the edge formed in the microdosimetric spectra collected at the Bragg peak and the distal fall-off region. A sigmoid curve is fitted to the edge region of which the intercept of the tangent through the inflection point with the horizontal axis is made to coincide with a specific lineal energy value. We calculated the maximum energy difference of particles traversing the SV for which the residual range difference in the continuous slowing down approximation (CSDA) is equal to the thickness of the sensitive volume of the microdosimeter. The microdosimetric spectra correspond to energy deposits in silicon, and the values of lineal energy are referred to silicon per unit of mass. This practice, although not generally adopted in microdosimetry, has important clinical advantages and derives from the correlation of linear energy with linear energy transfer (LET). The LET is the quantity which in the clinic is conventionally referred to the unit of mass. This is fundamental in treatment planning systems as the density of the matter crossed by the radiation varies significantly not only from tissue to tissue but, for the same tissue, from instantaneous conditions, as for example for lung tissue. Representations of the LET, and therefore of the lineal energy, which refer to the specific and instantaneous density of the target are not feasible, and the unit density is chosen as a reference. All the lineal energies reported in this work refer to silicon with unitary density and, in case one wants to obtain the values corresponding to the silicon density, it will be necessary to multiply the reported lineal energies by the factor 2.32. Because of the normalization to density, the conversion between materials becomes a mass stopping power rather than a stopping power ratio as is done in clinical dosimetry. The effects of angular deflection on the ion paths are considered negligible.

The frequency-mean lineal energy $\bar{y}_F$ and the dose-mean lineal energy $\bar{y}_D$ were estimated for the partial microdosimetric spectra corresponding to primary particles, the secondary fragments and the whole microdosimetric spectra. The $\bar{y}_D$ and the $\bar{y}_F$ are calculated considering that they are defined only in a certain interval above the noise cut-off value as follows[16]:

$$\bar{y}_{D\,(a,b)} = \int_{y_a}^{y_b} y \cdot d(y)\, dy \qquad (1)$$

$$\bar{y}_{F\,(a,b)} = \int_{y_a}^{y_b} y \cdot f(y)\, dy \qquad (2)$$

where $y_a$ and $y_b$ are the lower and higher lineal energy value in this interval and d(y) and f(y), respectively, are normalized over the interval. Although this does not corresponds to the values that would be obtained when using the entire spectrum, it allows to test the reproducibility and to compare the results of different microdosimeters if the cut-off energy is the same. Methods based on the extrapolation of the curve below the noise threshold have been adopted in other experiments. We decided here not to follow that procedure since the profiles of those extrapolations are inevitably arbitrary unless particle fluence and dose below the cutoff value are known.

While it would in principle be possible to detetermine the position of the detector in the holder mechanically, we deemed it more accurate to account for all sources of misalignment together by the following alignment procedureThe fluence per spot was estimated from the microdosimetric spectra by calculating the number of detected particles above the noise level divided by the number of repeated spots. The number of particles per spot is regulated by the beam delivery system and is considered constant throughout the test with a standard deviation of 1.5 %. The transversal distribution of the number of particles in a pencil beam is assumed to be, to first approximation, Gaussian with maximum values at the detector's center . They are assessed using the Gaussian function from the Python packages numpy and scipy to fit the experimental values of the fluence at the different lateral positions.

**Figure 3** shows the Gaussian distribution for one set of the data (position C), where the position where the laser corresponds to the fiducial references on the detector holder is chosen as zero and x is the displacement of the microdosimeter relative to that position. For this set of data, the mean was found to be $\mu_1$ = -2.2 mm. For the second set of measurements taken, $\mu_2$ = -1.4 mm was found and for the third set $\mu_3$ = -1.8 mm. To compensate for the displacement, the horizontal distance was adjusted for different sets of measurements taken over a period of a few months with an uncertainty of 0.4 mm calculated as a standard deviation of the three values. The vertical distance was obtained from one set of data and not calculated over different shifts and it was used for all the other sets of vertical lateral measurements. The mean offset for the vertical measurements was 1.2 mm. We take the same reproducibily uncertainty in the vertical direction as in the horizontal direction, thus the combined reproducibility uncertainty was found to be 0.6 mm. This uncertainty, even though it is poorly estimated, will not make a significant difference in the discussion as will become evident later.

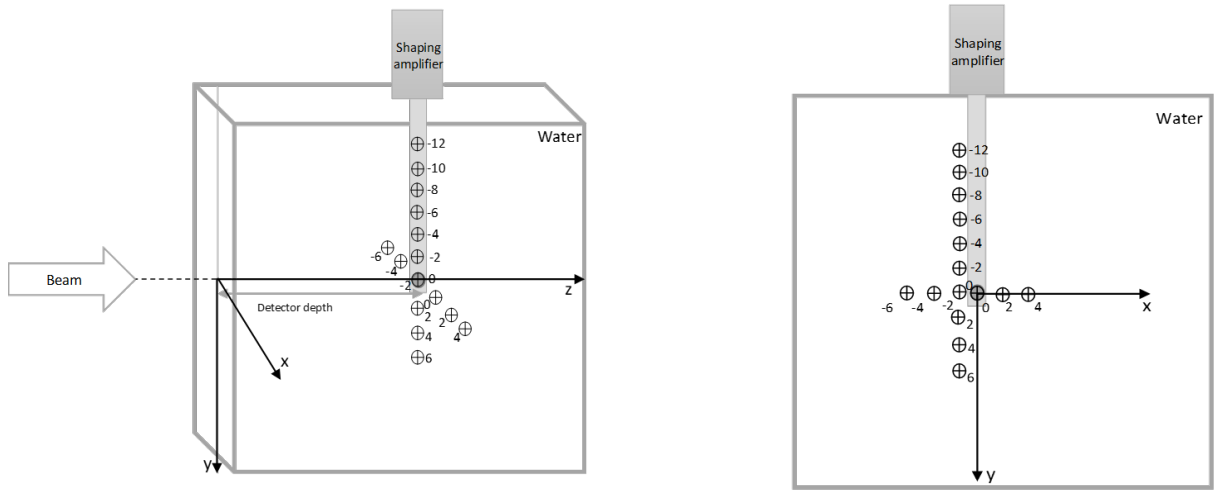

*Figure 1:* Schematic representation of the experimental set-up. Left picture showing the measuring positions on the vertical axis (y) and horizontal (lateral) axis (x) in the lateral view of the phantom with (z) representing the depth position; the right picture showing the measured points from the beam's eye view. The symbol ⊕ shows the positions of the detector during the measurements.

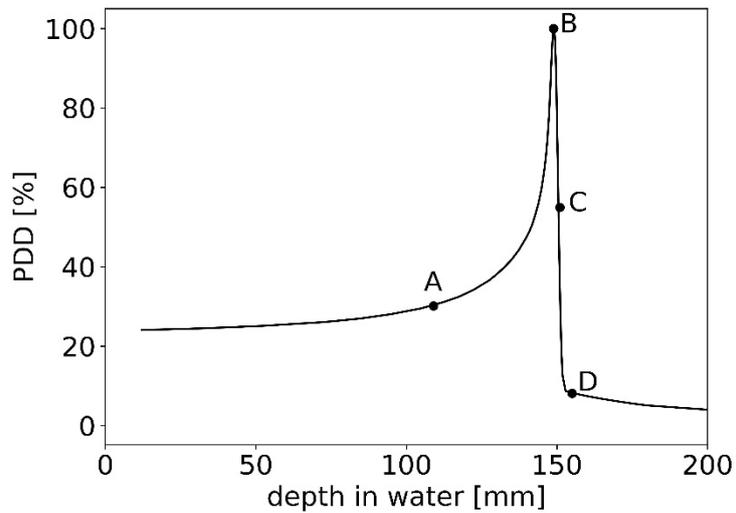

*Figure 2:* Relative integral depth dose profile for the carbon ion beam of nominally 284.7 MeV/u measured using PEAKfinder [14] and the positions along the z-axis for which the horizontal and vertical lateral measurements were taken.

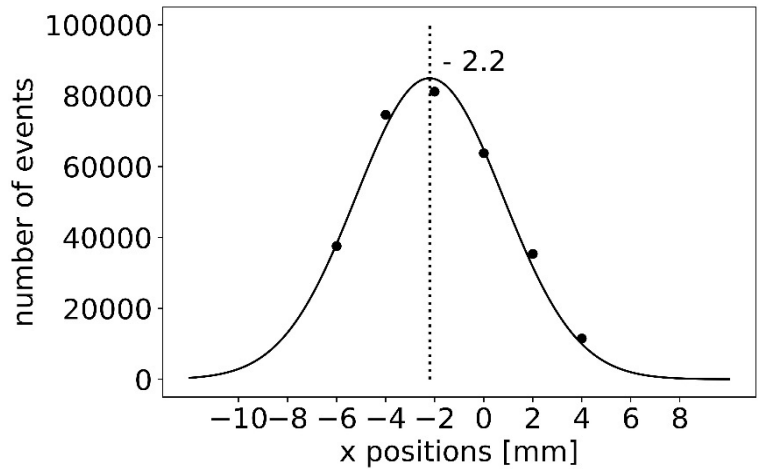

*Figure 3: Measured values (symbols) of the number of events for the different lateral positions relative to the central axis for the first set of data and the Gaussian fit to the data (curve)*

## 3. Results and discussion

The microdosimetric spectra in **Figure 4** show the spectra measured at the center and the two extreme horizontal positions. The spectra in the first few millimeters are not shown since they overlap with the central axis spectum, but from 5 mm, we start seeing an increase of a peak which can be attributed to the secondary fragments in the y range between 10 and 30 keV µm$^{-1}$.

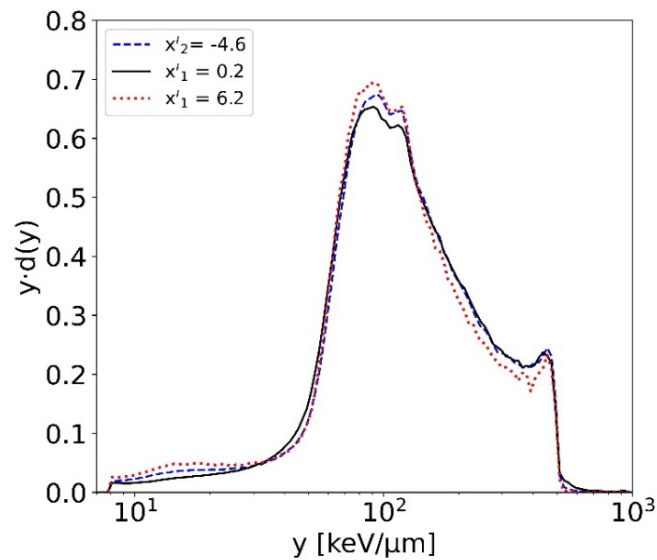

*Figure 4: yd(y) representation of the microdosimetric spectra measured at a depth of 15.1 cm at the center and two extreme horizontal positions*

The microdosimetric spectra measured in the vertical lateral positions at the four different depths are shown in **Figure 5**. As can be seen in panel (a) for the vertical variation of the measured spectra in the plateau (A), pile up is present as a small bump on the distal part of the microdosimetric spectra and as we go further away from the central axis the pile up is reduced. For y' = -11.2 mm and -13.2 mm the pile up disappears, and the peak of secondary fragments increases. In the Bragg Peak (B) shown in panel (b), the same behavior is also seen as we go away from the center.

The probability of pileup, depends on the transversal beam size and the particle flux. The beam size is constantly increasing with the penetration, The particle flux is affected by seasonal, inter-spill, and intra-spill variations. Historical data collected regularly at MedAustron facility show fluctuations as high as 30%.

Therefore, with the existing acquisition system, the *a priori* estimation of the pileup is not accurate. However for the most superficial acquisitions the pile up can be assessed *a posteriori* comparing the f(y) distributions at different distances from the beam axis.

More sophisticated acquisition systems may provide an estimation of the pile up recording each single waveform used in the MCA and analysing it to see if the profile is compatible, in risetime and decay time, with the pulses generated by a single particle.

In the fall-off (C) (panel (c)), the farther we are away from the center of the SV, the more the peak of the fragments is pronounced and the more the primary ions peak is reduced in the microdosimetric spectra. In the $R_{08}$ (D), we see some different behavior between the two microdosimeric spectra as, the spectrum at y' = -11.2 mm, is moved to lower LET. That could be a sign of a different percentage of low LET fragments, but given the low statistics, it is difficult to judge if the variation with off axis distance is significant.

The choice of the appropriate cross-sectional area of the sensitive volume is made by considering the flux and the transversal dimension of the beam. The size of the detector cross-sectional area is chosen based on two competing requirements, on the one hand the need to temporally separate the energy depositions from single particles and on the other the need to obtain spectra with good statistics. It is not always possible to reach a good compromise and this can result in spectra in which the signal pileup effect is evident. This is the case of the spectra collected near the beam axis (see figure). In the specific example described here, the drastic decrease in pileup due to fluence differences is shown when the sensitive volume is eccentric by a few millimeters. The experimental data reported here are useful for pileup correction studies in spectra. Considering that the other spectral characteristics are kept almost constant up to 4 mm from the beam axis, the spectra collected at different distances can provide the quantitative correlation between the fluence and the amount of distortion due to pileup.

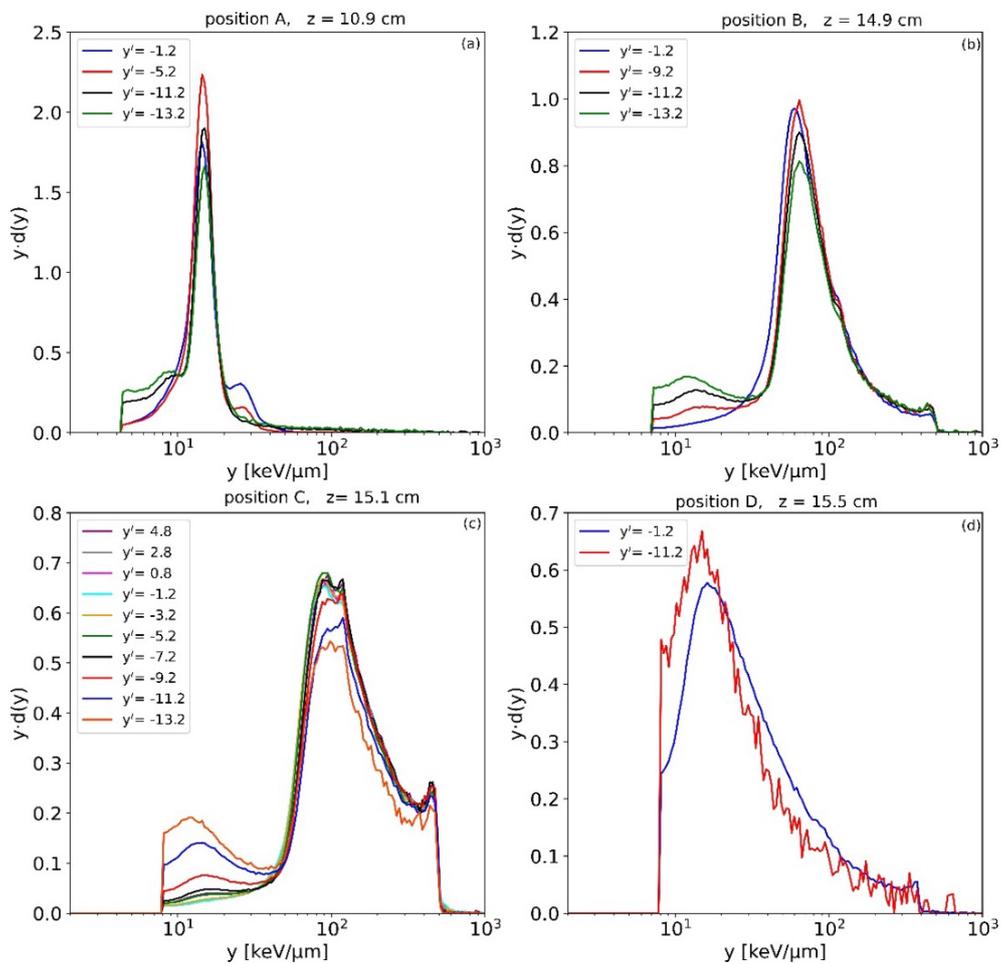

*Figure 5:* y·d(y) distribution at the four measured depth positions showing the vertical dependence of the measured spectra in (a) position A in the plateau, (b) position B in the Bragg Peak, (c) position C at $R_{55}$, (d) position D in the $R_{08}$

The values of $\bar{y}_{F(a,b)}$ and $\bar{y}_{D(a,b)}$ in the fall-off position (C) were calculated over three intervals, all channels above 8 keV.µm$^{-1}$, above 30 keV µm$^{-1}$ where the events are predominantly due to the primary particles and between 8 keV µm$^{-1}$ and 30 keV µm$^{-1}$ where we assume the events are predominantly due to the secondary fragments. 30 keV µm$^{-1}$ was chosen as an arbitrary cut-off value between the primary and secondary based on a phenomenological inspection of the experimental data. **Figure 6** represents the $\bar{y}_{F(a,b)}$ (left) and $\bar{y}_{D(a,b)}$ (right) and the relative 5 % uncertainty contribution coming from the ICRU stopping power data[15]. For the whole spectra, $\bar{y}_F$ varies between 4 % and up to 35 % for y = -13.2 mm and $\bar{y}_D$ varies between 1 % for y = -5.2 mm and 22 % for y = -3.2 mm. Above 30 keV.µm$^{-1}$, $\bar{y}_F$ and $\bar{y}_D$ varies between 0.1 % and 2 %. The variations in the sub-interval fall within the 5 % relative uncertainty contribution coming from the ICRU stopping power data. Although the primary and secondary components, each independently, are observed to have an almost constant average value, the overall $\bar{y}_{F(a,b)}$ and $\bar{y}_{D(a,b)}$ vary with the distance from the central axis because the dose fractions of the two components change. $\bar{y}_F$ depends more on the lateral distance than $\bar{y}_D$ because it is more sensitive to the lowest lineal-energy part of the distribution. If this could be parametrized as a function of depth and lateral position, this could give a great advantage in parametrization of $\bar{y}_F$ and $\bar{y}_D$ in pencil beam algorithms for treatment planning systems. The higher number of pulses piling up in the beam axis seems to have negligible effects on the determination of $\bar{y}_F$ and $\bar{y}_D$.

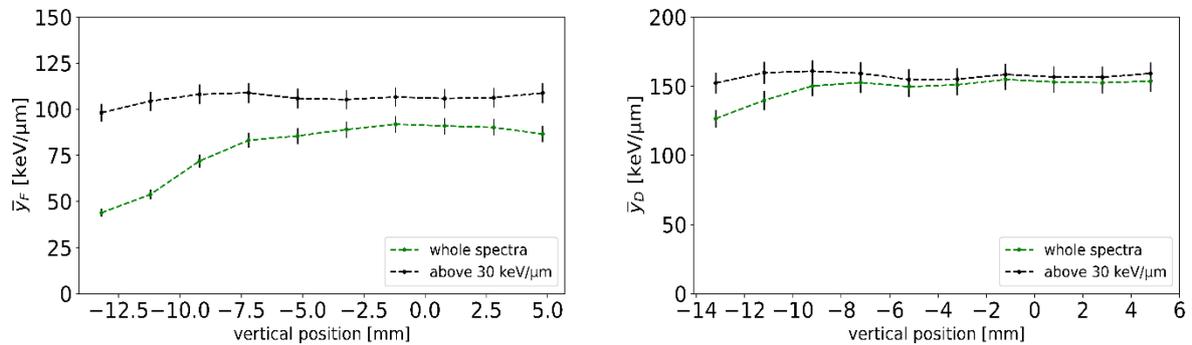

**Figure 6**: Left: $\bar{y}_F$ values for the overall spectra and for the predominantly primary particle fraction of the spectrum (above 30 keV µm$^{-1}$). Right: $\bar{y}_D$ values for the overall spectra and for the predominantly primary particle fraction of the spectrum (above 30 keV µm$^{-1}$).

Recent studies proposed parameters for the characterization of radiation in treatment planning based on the physical dose delivered locally below and above a suggested LET threshold of 30 keV·µm$^{-1}$ [18]. We can extend this indication from LET terms to lineal-energy terms, and subdivide the y·d(y) spectrum into the two components to the left and right of the value y = 30 keV·µm$^{-1}$. An example is shown in **Figure 7** which emphasizes the two components for the spectrum collected at depth B. The areas of the two components, shaded in blue and in green in the figure, are proportional to the respective doses imparted and it is evident how the two components vary at different distances from the axis.

The experimental results described here emphasize the importance of using a beam model that takes into account the transitions of the radiation quality at different distances from the beam axis.

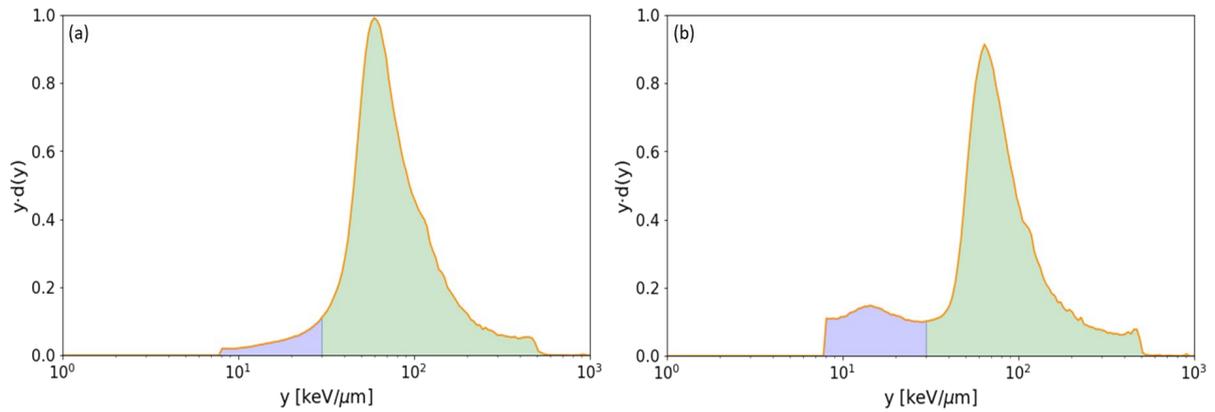

*Figure 7:* *Microdosimetric spectra collected at a depth of 14.9 cm and at the horizontal position -1.2 mm (a) and -13.2 mm (b); the blue and the green areas represent the dose below and above 30 keV.µm$^{-1}$ respectively. Although the spectra are incomplete at the lower lineal-energy values due to the cutoff of the noise, they show the evident variations of components at different distances from the beam axis which, potentially, may have clinical significance.*

The symmetry of the beam in both the vertical and horizontal positions is shown in **Figure 8**, with minor differences in the primary peak.

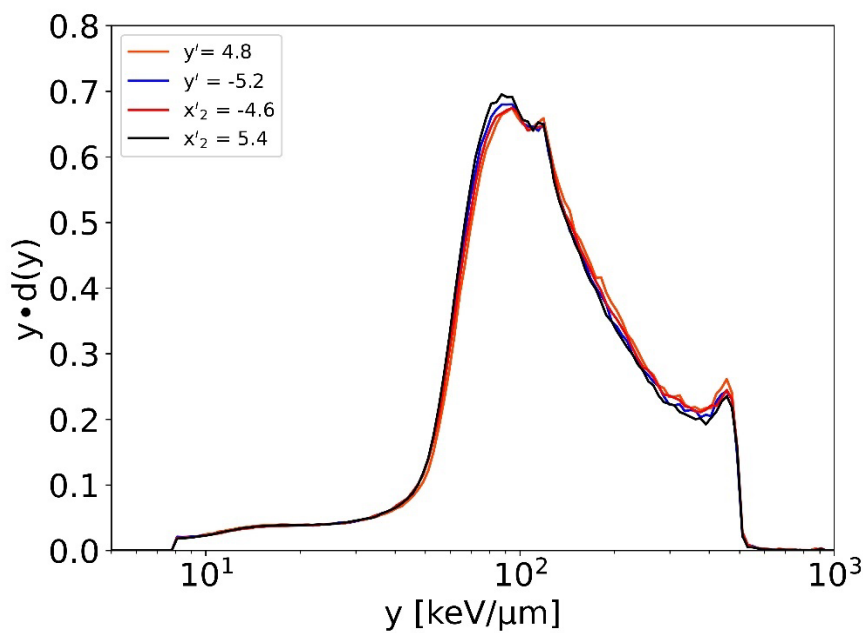

*Figure 8:* *Microdosimetric spectra horizontal and vertical symmetry at approximately 5 mm and at the depth of 15.1 cm*

In **Figure 9** the vertical position y' is fixed and the variation of the microdosimetric distribution with depth is shown. **Figure 9** (a) displays the microdosimetric spectra at y' = -1.2 mm, where the primary particle peak is dominant at the positions A, B and C. **Figure 9** (b) represents the microdosimetric spectra at the position y' = -11.2 mm. All four curves show an increased number of events in the lowest lineal energy range from the position y' = -1.2 mm, and with increasing depth, the secondary fragment peak increases and the primary peak decreases. At the beginning of the fragmentation tail, i.e., position D, the edge is moved towards a lower lineal energy (arrow II **Figure 9** (a)). Considering the dosimetric findings of Kempe *et al.*[19] we can assume that carbon ions are not present anymore. While a clear edge is formed, we see indication of Boron stoppers reaching the end of their range i.e., edge. In support of this hypothesis, the lineal energy at the edge was calculated from the stopping power for boron ions (**Table 1**). Comparing the ratio between the carbon- and boron-edge positions and the ratio of the lineal-energy edges predicted from the stopping power table, the values obtained were 1.29 and 1.26, respectively. This gives an indication of the presence of the boron at the edge. A study by Endo *et al.*[7] showed similar spectra where boron ions, identified through time-of-flight investigations, were the main contributor to the dose distribution at comparable depth conditions in the fragmentation tail; in that case in the edge region of the spectra and for the case 142.9 mm, Endo's data show significant fluctuations which seem to be related to low statistics and the presumed boron edge cannot be as clearly identified as in our work.

A Monte Carlo study was made by Barna et al.[20], where the same conditions were taken. The simulation were done for five position at y = -1.2 mm and y = -11.2 mm. The results indicates that the edge seen in the tail is indeed caused by boron ions.

A second bump is visible in the fall-off position around 117 keV µm$^{-1}$ (arrow I, **Figure 9** left), of which the relative amplitude increases with distance from the axis (arrow III, **Figure 9** right). The edge value was calculated considering the thickness of the detector according to the procedure described by Meouchi *et al.*[15] for the different ions up to carbon ions from the ICRU stopping power table (see **Table 1**). Based on this we hypothesize that this bump is associated with helium ion stoppers.

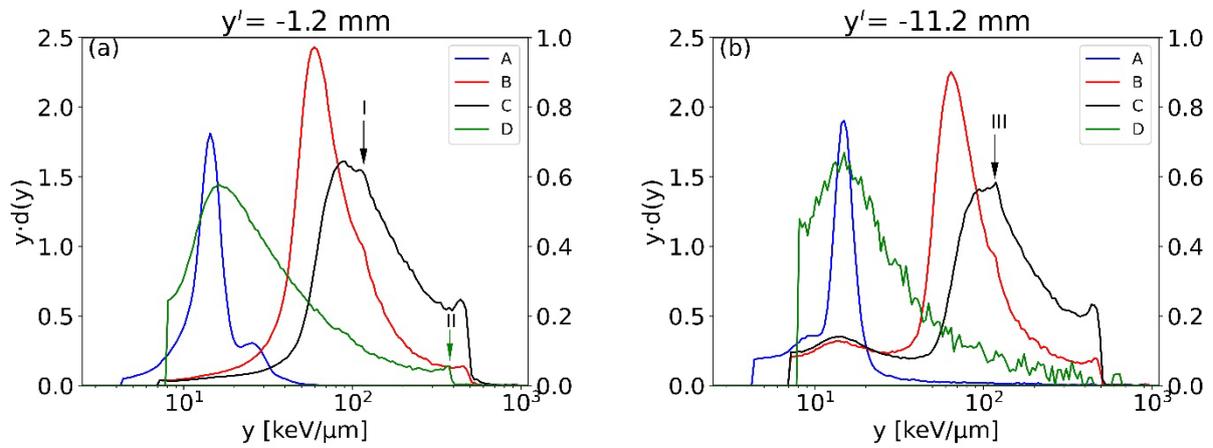

***Figure 9***: *y·d(y) distribution, in the plateau (A), at the Bragg peak (B), the distal fall-off (C), and in the fragmentation tail (D). (a) with the center of the spot coinciding with the center of the SV, i.e., y' = -1.2 mm, (b) at the vertical position y' = -11.2 mm*

***Table 1:*** *the maximum lineal-energy edge for the different ions calculated from the stopping power table ICRU and considering the dependence on the thickness of the microdosimeter*[15]

| Ions | Stopping power (keV μm$^{-1}$) |
|------|-------------------------------|
| p    | 30.3 ± 1.5                    |
| He   | 113.3 ± 5.7                   |
| Li   | 203.0 ± 10.2                  |
| Be   | 301.04 ± 15.1                 |
| B    | 405.3 ± 20.3                  |
| C    | 511.2 ± 25.6                  |

## 4. Conclusion

This investigation on carbon ions shows the relevance of comparing microdosimetric measurements in-axis and out-of-axis to find the microdosimetric spectra characteristics associated to specific fragments, that would be otherwise, not be recognized. The 3D SOI mushroom silicon microdosimeter was used to characterize the 284.7 MeV/u carbon-ion pencil beam at the MedAustron ion therapy center. It was shown that up to 4 mm away from the beam's central axis, the microdosimetric spectra do not show noticeable variation. Starting from 5 to 6 mm lateral displacement, the lower lineal energy part of the microdosimetric spectrum shows a second peak that is growing as we move away from the central axis due to the increasing partial contribution of secondary fragments.

Considering two distinct parts of the microdosimetric spectra as pertaining to secondary and primary particle components, we found that each component separately exhibited partial microdosimetric spectra and mean values $\bar{y}_F$ and $\bar{y}_D$ that were almost invariant of lateral position. However, the relative dose contribution of each component changes with the lateral position. These features could be useful in the pencil beam algorithm for use in treatment planning system taking into account radiation quality of the field produced by pencil carbon ion scanning beam, under the condition that this observation holds and is verified for a wider range of conditions (initial beam energy, depth) as explored in this paper. This affects the total spectra and their mean values such that the $\bar{y}_F$ and $\bar{y}_D$ values decrease with increasing lateral offset due to the increasing weight of the secondary particle contribution. This has the largest effect on the frequency mean values $\bar{y}_F$, with a variation of up to 30 % because of their sensitivity to the lower lineal-energy contributions. The dose mean values $\bar{y}_D$ are less sensitive to the lower lineal-energy part and show a variation of up to 10 %.

Another effect was observed with increasing distance from the central beam axis, the increase of small additional bumps in the peak of the primary particles. Increasing the depth in water, a new edge can be seen in the contamination tail. Therefore, at a distance from and along the beam axis, microdosimetry potentially allows the distinction of different ion species, such as helium and boron fragments.

The studies described here characterize the ion beam primary and secondary particles substructure with radiobiologically relevant parameters, which are complementary to the dosimetry and provide an indication of the radiation quality. Based on these outcomes, the beam models, used at the facility in TPS and in radiobiological studies, can undergo additional validation and, if necessary, adjustment with an increased number of measurable parameters if the microdosimetric parameters are included in the beam commissioning process.

## Acknowledgements

The authors would like to thank Markus Stock for his valuable contribution in facilitating both the study and the measurements. They are also grateful for Rita Viegas and Yasmin Hamad for their time and help during the measurements. The work of Sandra Barna in this research was funded by the Austrian Science Fund (FWF) through project number P32103-B.

## Data availability statement

The data that support the findings of this study are available upon request from the authors.